%% file: main.tex
\documentclass[journal]{IEEEtran}
\usepackage{tikz}
\usepackage{cite}
\usepackage{multirow}
\usepackage{amsmath}
\DeclareMathOperator*{\argmax}{arg\,max}
\usepackage{amssymb}
\usepackage{graphicx}
\usepackage{hyperref}
\usetikzlibrary{positioning}
\usepackage{etoolbox}
\makeatletter
\patchcmd{\@makecaption}
  {\scshape}
  {}
  {}
  {}
\makeatletter
\usepackage{arydshln}

\begin{document}

\title{Designing Neural Speaker Embeddings \\ with Meta Learning}

\author{Manoj~Kumar,~\IEEEmembership{Member,~IEEE,}
        Tae Jin-Park,~\IEEEmembership{Member,~IEEE,}
        Somer~Bishop,~\IEEEmembership{}
        and~Shrikanth~Narayanan,~\IEEEmembership{Fellow,~IEEE}
\thanks{M. Kumar, T. J. Park and S. Narayanan are with Signal Analysis and Interpretation Laboratory, University of Southern California, Los Angeles, USA e-mail: (prabakar@usc.edu;taejinpa@usc.edu;shri@ee.usc.edu).
S. Bishop is with Department of Psychiatry, University of California, San
Francisco, USA e-mail:(somer.bishop@ucsf.edu)}
}


\maketitle

\begin{abstract}
Neural speaker embeddings trained using classification objectives have demonstrated state-of-the-art performance in multiple applications. 
Typically, such embeddings are trained on an out-of-domain corpus on a single task e.g., speaker classification, albeit with a large number of classes (speakers). 
In this work, we reformulate embedding training under the meta-learning paradigm. 
We redistribute the training corpus as an ensemble of multiple related speaker classification tasks, and learn a representation that generalizes better to unseen speakers.
First, we develop an open source toolkit to train x-vectors that is matched in performance with pre-trained Kaldi models for speaker diarization and speaker verification applications. We find that different bottleneck layers in the architecture variedly favor different applications. 
Next, we use two meta-learning strategies, namely prototypical networks and relation networks, to improve over the x-vector embeddings. Our best performing model achieves a relative improvement of 12.37\% and 7.11\% in speaker error on the DIHARD II development corpus and the AMI meeting corpus, respectively. We analyze improvements across different domains in the DIHARD corpus. Notably, on the challenging child speech domain, we study the relation between child age and the diarization performance.
Further, we show reductions in equal error rate for speaker verification on the SITW corpus (7.68\%) and the VOiCES challenge corpus (8.78\%). 
We observe that meta-learning particularly offers benefits in challenging acoustic conditions and recording setups encountered in these corpora.
Our experiments illustrate the applicability of meta-learning as a generalized learning paradigm for
training deep neural speaker embeddings.
\end{abstract}


\IEEEpeerreviewmaketitle

\input{sections/intro}

\input{sections/background}

\input{sections/methods}

\input{sections/datasets}

\input{sections/experiments}

\input{sections/conclusion}



\bibliographystyle{IEEEtran}
\bibliography{mybib}


\end{document}

%% file: sections/intro.tex
\section{Introduction}
\label{sec:intro}


Audio speaker embeddings refer to fixed-dimensional vector representations extracted from variable duration audio utterances and assumed to contain information relevant to speaker characteristics. In the last decade, speaker embeddings have emerged as the most common representations used for speaker-identity relevant tasks such as speaker diarization (speaker segmentation followed by clustering: \textit{who spoke when?}) \cite{anguera_DiarOverview2012} and speaker verification \textit{(does an utterance pair belong to same speaker?}) \cite{campbell_speakerRecogTutorial1997}.
Such applications are relevant across a variety of domains such as 
voice bio-metrics \cite{rahulamathavan_bioMetrics2019, scheffer_bioMetrics2013}, automated meeting analysis \cite{anguera2007acoustic,vanLeeuwen_meeting2006}, and clinical interaction analysis \cite{pal_clusterGAN2020, xiao2016technology}. Recent technology evaluation challenges \cite{ryant2019second, richey2018voices, Hansen_fearlessSteps2018, McLaren_SITW2016} have drawn attention to these domains by incorporating natural and simulated in-the-wild speech corpora exemplifying the many diverse technical facets that need to be addressed. 

While initial efforts toward training speaker embeddings had focused on generative modeling \cite{reynolds2000speaker,campbell_SVMGMM2006} and factor analysis \cite{dehak_ivectors2011}, deep neural network (DNN) representations extracted at bottleneck layers have become the standard choice in recent works. The most widely used representations are trained using a classification loss (d-vectors \cite{origdvec_variani2014deep}, x-vectors \cite{snyder_xvec2017, snyder_xvec2018}), while other training objectives such as triplet loss \cite{bredin_tristounet2017, zhang_triplet2018} and contrastive loss \cite{chung2018Voxceleb2} have also been explored.
More recently, end-to-end training strategies \cite{Fujita2019, horiguchi2020endtoend, fujita2020endtoend} have been proposed for speaker diarization to address the mismatch between training objective (classification) and test setup (clustering, speaker selection, etc).

A common factor in the classification formulation is that all the speakers from training corpora are used throughout the training process for the purpose of loss computation and minimization. Typically, categorical cross-entropy is used as the loss function.
While the number of speakers (classes) can often be large in practice ($\mathcal{O}(10^3)$), the classification objective represents a single task, i.e., the same speaker set is used to minimize cross-entropy at every training minibatch.
This entails limited task diversity during the training process and offers scope for training better speaker-discriminative embeddings by introducing more tasks.
We note that a few approaches exist which introduce multiple objectives for embedding training, such as metric-learning with cross entropy \cite{XU2020394, ren2019} and speaker classification with domain adversarial learning \cite{zhou2019dann, wang2018dann}. While these approaches demonstrate improvements over a single training objective, the speaker set is often common across objectives (except in domain adversarial training where target speaker labels are assumed unavailable).

In this work we use the classification framework while training neural speaker embeddings, however we decompose the original classification task into multiple tasks wherein each training step optimizes on a new task. 
A common encoder is learnt over this ensemble of tasks and used for extracting speaker embeddings during inference.
At each step of speaker embedding training, we construct a new task by sampling speakers from the training corpus. For a large training speaker set available in typical training corpora,
generating speaker subsets results in a large number of tasks.
This provides a natural regularization to prevent task over-fitting. 
Our approach is inspired by the meta-learning \cite{Schmidhuber_thesis} paradigm, also known as {\it learning to learn}. Meta-learning optimizes at two-levels: within each task and across a distribution of tasks \cite{ravi2017}. This is in contrast to conventional supervised learning which optimizes a single task over a distribution of samples. 
In addition to benefits from increased task variability meta-learning has demonstrated success in unseen classes \cite{ravi2017, finn_maml2017,Andrychowicz_2016}.
This forms a natural fit for applications such as speaker diarization and speaker verification which often evaluate on speakers unseen during embedding training.

We compare our meta-learned models with x-vectors, which have established state-of-the-art performance in multiple applications \cite{snyder_xvec2017, snyder_xvec2018} including recent evaluation challenges such as DIHARD\cite{Sell2018_dihard} and VOiCES \cite{richey2018voices}. 
First, we develop a competitive wide-band x-vector baseline using the PyTorch toolkit (calibrated with identical performance with the Kaldi Voxceleb recipe\footnote{https://github.com/kaldi-asr/kaldi/tree/master/egs/voxceleb}). 
Next, we use two different metric-learning objectives to meta-learn the speaker embeddings: prototypical networks and relation networks. While both approaches share the task sampling strategy during the training phase, they differ in the choice of the comparison metric between samples. We evaluate our approaches on two different applications: speaker diarization and speaker verification to illustrate the generalized speaker discriminability nature of meta-learned embeddings. 

The contributions of this work are as follows: we develop new speaker embeddings using meta-learning that are not restricted to an application.
Within each application, we demonstrate improvements using multiple corpora obtained under controlled as well as naturalistic speech interaction settings.
Furthermore, we identify conditions where meta-learning demonstrates benefits over conventional cross-entropy paradigm. 
We analyze diarization performance across different domains in the DIHARD corpora. We also consider the special case of impact of child age groups using internal child-adult interaction corpora from the Autism domain. We study the effect of data collection setups (near-field, far-field and obstructed microphones) and the level of degradation artifacts on the speaker verification performance.
While we present results using prototypical networks and relation networks, the proposed framework is independent of the specific metric-learning approach and hence offers scope for incorporating non-classification objectives such as clustering. It should be noted however that the application of relation networks has not been explored in speaker embedding research.
Finally, we present an open source implementation of our work, including x-vectors baselines, based on a generic machine learning toolkit (PyTorch)\footnote{https://github.com/manojpamk/pytorch\_xvectors}.

%% file: sections/background.tex
\section{Background}
\label{sec:background}



\subsection{Meta-Learning for Task Generalization}

Early works on meta-learning focused on adaptive learning strategies such as combining gradient descent with evolutionary algorithms \cite{yao_evolve1999,ABRAHAM20041}, learning gradient updates using a meta-network \cite{naik_metaNN1992} and using biologically inspired constraints for gradient descent \cite{bengio_synaptic1991, bengio1992optimization}.
Recent meta-learning approaches have addressed the issue of rapid generalization in deep learning, by learning to learn for a new task \cite{Andrychowicz_2016, finn_maml2017, ravi2017}.
This concept is inspired by the human ability to learn using a handful of examples. For instance children learn to recognize a new animal when presented with a few images as opposed to conventional DNNs which require thousands of samples for a new class.
The ability to quickly generalize to unseen classes is achieved by generating diversity in training tasks, for instance by using different sets of classes at each training step (see Fig. 1 in \cite{ravi2017}). Further, the classification setup (in terms of number of classes and samples per class) is controlled to match with that of the test task \cite{snell2017prototypical}.
Meta-learning has been successfully applied to achieve task generalization in computer vision \cite{ravi2017,snell2017prototypical,finn_maml2017} and more recently in natural language processing \cite{yu2018diverse,GaoH0S19, dou-etal-2019-investigating}.
Drawing parallels with the above applications, we train speaker embeddings with a large number of speaker classification tasks to improve over the conventional model which uses a single classification task. Since speaker sets differ between training steps, we replace the conventional softmax nonlinearity and cross-entropy loss combination with metric learning objectives used in previous meta-learning works \cite{snell2017prototypical,sung2018learning,vinyals2016matching,geng2019induction}.

\subsection{Meta-Learning Speaker Embeddings}

Few recent approaches have used a variant of meta-learning to train speaker embeddings, specifically the metric-learning objective from prototypical networks (protonets).
In \cite{chung2020defence}, the authors extend angular softmax objective to protonets and compare with various metric learning approaches for speaker verification. Across different architectures, angular prototypical loss outperforms other methods including conventional softmax objective. 
The authors in \cite{kye2020metalearning} applied protonets for short utterance speaker recognition and introduced global prototypes that mitigate the need for class sampling. 
In related applications, \cite{ko_protonets2020} and \cite{an2019shot} used protonets for small footprint speaker verification and few-shot speaker classification, respectively.
In \cite{wang_centroid2019}, the protonet loss was compared with triplet loss and evaluated on (open and close set) speaker ID and speaker verification tasks. 
However, previous approaches seldom compare embeddings trained using protonets with existing benchmarks based on x-vectors, except for \cite{ko_protonets2020} where a modified architecture was used owing to the nature of the task. Further, the class sampling strategy is not always used with protonets (e.g., \cite{chung2020defence,kye2020metalearning})
which might inhibit task diversity during training. 
An exception from the above metric-learning approaches is \cite{kang2020domaininvariant}, where the authors train deep speaker embeddings using the model-agnostic meta-learning strategy to mitigate  domain mismatch for speaker verification.
To the best of our knowledge, meta-learning is yet to be applied for general-purpose speaker diarization, except for the specific case of dyadic speaker clustering in child-adult interactions in our recent work \cite{koluguri2020}. 


%% file: sections/methods.tex
\section{Methods}
\label{sec:methods}

In this section, we introduce the meta-learning setup for neural embedding training followed by description of two metric-learning approaches adopted in this work: prototypical networks and relation networks. Following which, we outline their use in our tasks: speaker diarization and speaker verification, including a description of the choice of clustering algorithm.

Consider a training corpus where $C$ denotes the set of unique speakers, and where each speaker has multiple utterances available. Typically, $|C|$ is a large integer
($\mathcal{O}(10^3)$).  
Here, an utterance might be in the form of raw waveform or frame-level features such as MFCCs or Mel spectrogram. 
Under the meta-learning setup, each episode (a training step; equivalent to a minibatch) consists of two stages of sampling: classes and utterances conditioned on classes. 
First, a subset of classes $L$ (speakers) is sampled from $C$ within an episode, with the number of speakers per episode $|L|$ typically held constant during the training process. 
Next, two disjoint sets from each speaker in $L$ are sampled without replacement from the set of all utterances belonging to that speaker: supports $S$ and queries $Q$. 
Within an episode, supports and queries are used for model training and loss computation, respectively, similar to train and test sets in supervised training. This process continues across a large number of episodes with speakers and utterances sampled as explained above.
Following terminology from Section \ref{sec:intro}, an episode is equivalent to a \textit{task}, wherein the model learns to classify speakers from that task. Hence, meta-learning optimizes across tasks, treating each task as a training example. The optimization process is given as:
\begin{equation}
    \theta = \argmax_{\theta} \displaystyle \mathop{\mathbb{E}}_{L} [ \mathop{\mathbb{E}}_{S, Q} [ \mathop{\mathbb{E}}_{(\mathbf{x},y) \in Q} [ \log p_\theta (y|\mathbf{x},S)] ]]
\end{equation}
Here, $\theta$ denotes trainable parameters of the neural network, $(\mathbf{x},y)$ represents an utterance and its corresponding speaker label. In contrast to conventional supervised learning:
\begin{equation}
    \theta = \argmax_{\theta} \displaystyle \mathop{\mathbb{E}}_{B} [  \mathop{\mathbb{E}}_{(\mathbf{x},y) \in B}[\log p_\theta (y|\mathbf{x})]  ]
\end{equation}
where $B$ denotes a minibatch. Meta-learning approaches are broadly categorized based on the characterization of $p_\theta (y|x)$: model-based \cite{santoro2016meta}, metric-based \cite{vinyals2016matching} and optimization-based meta-learning \cite{finn_maml2017}. 
Of interest in this work are metric-based approaches where $p_\theta (y|x)$ is a potentially learnable kernel function between utterances from $S$ and $Q$. The reasoning is as follows: speaker embeddings trained for classification are bottleneck representations, and the latter is directly optimized using task performance in metric-learning approaches. We now describe the two metric-learning approaches used in this work: prototypical networks and relation networks.

\subsection{Prototypical Networks}

Protonets learn a non-linear transformation where each class is represented by a single point in the embedding space, namely the centroid (prototype) of training utterances from that class. During inference a test sample is assigned to the class of nearest centroid, similar to the nearest class mean method \cite{mensink2013}. 

At training time, consider an episode $t$, the support set ($S_t$) and the query set ($Q_t$) sampled as explained above. Supports are used for prototype computation while queries are used for estimating class posteriors and loss value. 
The prototype ($\mathbf{v}_c$) for each class is computed as follows:
\begin{equation}
\label{eqn:proto_basic}
    \mathbf{v}_{c} = \frac{1}{|S_{t,c}|} \sum_{(\mathbf{x_{i}},y_{i})\in S_{t,c}} f_{\theta}(\mathbf{x}_i)
\end{equation}
$f_{\theta} : \mathbb{R}^M \rightarrow \mathbb{R}^P$ represents the parameters of the protonet. $\mathbf{x_i}$ represents an $M$-dimensional utterance representation extracted using a DNN.
$S_{t,c}$ is the set of all utterances in $S_t$ belonging to class $c$. For every test utterance $\mathbf{x_j} \in Q_t$, the posterior probability is computed by applying softmax activation over the negative distances with prototypes:
\begin{equation}
\label{sofmax-eq2}
    p_\theta(y_j=c\,| \mathbf{x}_j, S_t)=\frac{\exp \left(-d\left(f_{\theta}(\mathbf{x}_j), \mathbf{v}_c\right)\right)}{\sum_{c^{\prime} \in L} \exp \left(-d\left(f_{\theta}(\mathbf{x}_j), \mathbf{v}_{c^{\prime}}\right)\right)}
\end{equation}
$d$ represents the distance function. 
Squared Euclidean distance was proposed in the original formulation \cite{snell2017prototypical} due to its interpretability as 
a Bregman divergence \cite{banerjee2005} as well as supporting empirical results. For the above reasons, we adopt squared Euclidean as a metric in this work.
The negative log-posterior is treated as the episodic loss function and minimized using gradient descent:
\begin{equation}
\label{eqn:proto_backprop}
 \text{Loss} = - \frac{1}{|Q_t|} \sum_{(\mathbf{x}_j,y_{j})\epsilon Q_t } \log(p_\theta(y_j \mid \mathbf{x}_j, S_t)) 
\end{equation}

\subsection{Relation Networks}

Relation networks compare supports and queries by learning the kernel function simultaneously with the embedding space\cite{sung2018learning}. In contrast with protonets which use squared Euclidean distance, relation networks learn a more complex inductive bias by parameterizing the comparison metric using a neural network. 
Hence, relation networks attempt to jointly learn the embedding and metric over an ensemble of tasks that are generalized to an unseen task. Specifically, there exist two modules: an encoder network that maps utterances into fixed-dimensional embeddings and a comparison network that computes a scalar relation given pairs of embeddings. Given supports $S_t$ within an episode $t$, the class representation is taken as the sum of all support embeddings:
\begin{equation}
\label{eqn:relation_encoder}
    \mathbf{v}_c = \sum_{(\mathbf{x}_i,y_{i})\epsilon S_{t,c}} f_{\theta}(\mathbf{x}_i)
\end{equation}
$f_{\theta}$ represents the encoder network. For each query embedding belonging to a class $j$, its relation score $r_{c,j}$ with training class $c$ is computed using the comparison network $g_{\phi}$ as follows:
\begin{equation}
\label{eqn:relation_relation}
    r_{c,j} = g_{\phi} ([ \mathbf{v}_c, f_{\theta} (\mathbf{x}_j) ])
\end{equation}
Here $[.,.]$ represents concatenation operation. The original formulation of relation networks \cite{sung2018learning} treated the relation score as a similarity measure, hence $r_{c,j}$ is trained with:

\begin{equation}
    r_{c,j}= 
\begin{cases}
    1, & \text{if } y_j = c\\
    0, & \text{otherwise}
\end{cases}
\end{equation}

In the original formulation \cite{sung2018learning}, the networks $f_\theta$ and $g_\phi$ were jointly optimized using mean squared error (MSE) objective since the predicted relation network was treated similar to a linear regression model output. In this work, we replace MSE with the conventional cross-entropy objective based on empirical results. Hence the posterior probability is computed as:
\begin{equation}
\label{softmax-eq}
    p_\theta(y_j| \mathbf{x_j}, S_t)=\frac{\exp \left(  r_{c,j}   \right)}{\sum_{c^{\prime} \in L} \exp \left(    r_{c^{\prime},j}     \right)}
\end{equation}
and the loss function is computed using Eq. \ref{eqn:proto_backprop}.

\subsection{Use in Speaker Applications}

\subsubsection{Speaker Diarization}
\label{subsubsec:bSCNME}
Typically, there exist four steps in a speaker diarization system: speech activity detection, speaker segmentation, embedding extraction and speaker clustering (exceptions include recently proposed end-to-end approaches \cite{horiguchi2020endtoend,fujita2020endtoend}). In this work, we adopt the uniform segmentation strategy similar to \cite{Sell2018_dihard, garciaRomero2017} wherein the session is segmented into equal duration segments with overlap. Meta-learned embeddings are extracted from these segments followed by clustering. We use a recently proposed variant of spectral clustering \cite{park_SC2020} which uses a binarized version of affinity matrix between speaker embeddings. The binarization is expressed using a parameter ($p$) which represents the fraction of non-zero values at every row in the affinity matrix.
The clustering algorithm attempts a tradeoff between pruning excessive connections in the affinity matrix (minimizing $p$) while increasing the normalized maximum eigengap (NME; $g_p$) where the latter is expressed as a function of $p$ (Eq. (10) in \cite{park_SC2020}). The ratio ($\frac{p}{g_p}$) is then minimized to estimate the number of resulting clusters (i.e., speakers) in a session. This process is referred to as binarized spectral clustering with normalized maximum eigengap (NME-SC).

Our choice of NME-SC in this work is motivated by two reasons: (1) We do not require a separate development set to estimate a threshold parameter used in the more common agglomerative hierarchical clustering (AHC) method with average linking applied on distances estimated using probabilistic linear discriminant analysis (PLDA) \cite{Sell2018_dihard}. We choose the binarization parameter ($p$) for each session by optimizing for ($\frac{p}{g_p}$) over a pre-determined range for $p$. (2) Empirical results which demonstrate similar performance between AHC tuned on a development set and NME-SC reported in \cite{park_SC2020} and in this work. 

\subsubsection{Speaker Verification}
We use the standard protocol for speaker verification wherein a speaker embedding is extracted from the entire utterance. Subsequently, the embeddings are reduced in dimension using LDA and trial pairs are scored using a PLDA model trained on the same data used to train embeddings. Following this, target/imposter pairs are determined using a threshold on the PLDA scores.

\begin{figure*}
    \centering
    \includegraphics[width=\textwidth]{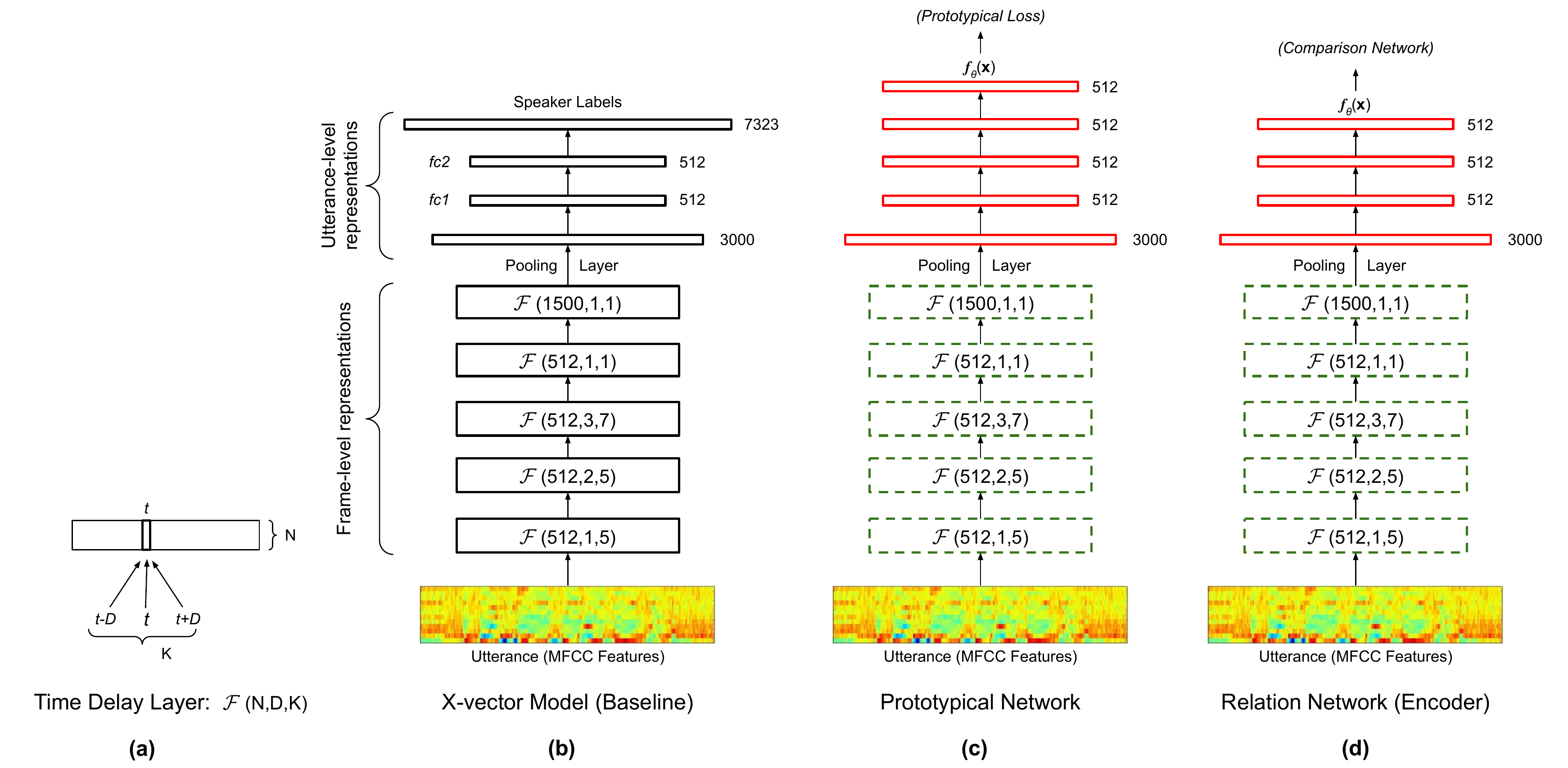}
    \caption{Overview of baseline and meta-learning architectures. \textbf{(a)} A time-delay layer $\mathcal{F}(N,D,K)$ which forms the basic component across models. At each time-step, activations from the previous layer are computed using a context width of $K$ and a dilation of $D$. $N$ represents the output embedding dimension. \textbf{(b)} Baseline x-vector model. Kaldi speaker embeddings are extracted at fc1 layer. We find that fc2 and fc1 embeddings perform better for speaker diarization and speaker verification respectively. \textbf{(c)} Prototypical network architecture. Layers marked with a dashed boundary are initialized with pre-trained x-vector models, while layers with a solid boundary are randomly initialized. The final layer output is referred to as protonet embeddings. \textbf{(d)} Relation encoder architecture. The final layer output is referred to as relation network embeddings. Relation scores are computed used these embeddings as illustrated in Fig. \ref{fig:metaLearningLoss}b) }
    \label{fig:metaLearningArch}
\end{figure*}



%% file: sections/datasets.tex
\section{Datasets}
\label{sec:dataset}

Since we evaluate meta-learned embeddings on two applications: speaker diarization and speaker verification, we use different corpora commonly used in evaluating these respective applications. We choose corpora obtained from both controlled and naturalisitc settings, with the former generally assumed relatively free from noise, reverberation and babble. We further choose additional corpora to assist with application-specific analysis of performance, such as the effect of domains and speaker characteristics (age) on diarization error rate (DER) and channel conditions on equal error rate (EER). A summary of the corpora used in this work is presented in Table~\ref{tab:corpora_overview}. Below, we provide details for each corpora.

\begin{table}[h]
\caption{Overview of training and evaluation corpora}
\label{tab:corpora_overview}
\centering{
\begin{tabular}{ccc} \\ \hline
\multicolumn{1}{c}{\multirow{2}{*}{\textbf{Training}}} & \multicolumn{2}{c}{\textbf{Evaluation}} \\ 
\multicolumn{1}{c}{} & Speaker Diarization & Speaker Verification \\ \hline
\rule{0pt}{2ex} Vox2 & AMI & Vox1 test \\
Vox1 dev & DIHARD II dev & VOiCES \\
 & ADOS-Mod3 & SITW \\ \hline
\end{tabular}}
\end{table}

\subsection{Voxceleb}
\label{subsec:data_vox}
The Voxceleb corpus \cite{chung2018Voxceleb2} consists of YouTube videos and audio of speech from celebrities with a balanced gender distribution. Over a million utterances from $\approx$7300 speakers are annotated with speaker labels. The utterances are collected from varied background conditions to simulate an in-the-wild collection. The Voxceleb corpus is further subdivided into Vox1 and Vox2 datasets. Following the baseline Kaldi recipe\footnote{https://github.com/kaldi-asr/kaldi/tree/master/egs/voxceleb/v2}, we use the dev and test splits from Vox2 and the dev split from Vox1 for embedding training. The test split from Vox1 is reserved for speaker verification. There exists no speaker overlap between the train set and Vox1-test set.

\subsection{VOICES}
The VOiCES corpora \cite{richey2018voices} was released as part of the VOiCES from a distance challenge\footnote{https://voices18.github.io/}. It consists of clean audio (Librispeech corpus\cite{panayotov_librispeech2015}) played inside multiple room configurations and recorded with microphones of different types and placed at different locations in the room. In addition, various distractor noise signals were played along with the source audio to simulate acoustically challenging conditions for speaker and speech recognition. Furthermore, the audio source was rotated in its position to simulate a real person. 
We use the evaluation portion of the corpus which is expected to contain more challenging room configurations \cite{evalplan_2019voices} than the development portion.

\subsection{SITW}
The speakers-in-the-wild corpus \cite{McLaren_SITWCorpus2016} was released as part of the SITW speaker recognition challenge. It consists of in-the-wild audio collected from a diverse range of recording and background conditions. In addition to speaker identities, the utterances are manually annotated for gender, extent of degradation, microphone type and other noise conditions in order to aid analysis. A subset of the utterances also include multiple speakers, with timing information available for the speaker with longest duration. 
A handful of speakers from the SITW corpus are known to overlap with the Voxceleb corpus~\footnote{\url{http://www.robots.ox.ac.uk/\~vgg/data/Voxceleb/SITW_overlap.txt}}. In this work, we remove the utterances corresponding to these speakers before evaluation. Details of corpora used in speaker verification is provided in Table~\ref{tab:spkrVerDataStats}.

\begin{table}[]
\caption{Statistics of corpora used for speaker verification, including trial subsets created for analysis purposes}
\label{tab:spkrVerDataStats}
\centering{
\begin{tabular}{llll} \\ \hline

\multicolumn{1}{c}{\textbf{Corpus}} & \multicolumn{1}{c}{\textbf{\#Spkrs}} & \multicolumn{1}{c}{\textbf{\#Utterances}} &
\textbf{\#Trails (\#target)} \\ \hline
Vox1 test & 40 & 4715 & 38K (19K) \\
VOiCES & 100 & 11392 & 3.6M (36K) \\
\quad close mic & 98 & 1076 & 0.84M (8.5K)\\
\quad far mic & 96 & 1006 & 0.78M (7.9K)\\
\quad obs mic & 96 & 1006 & 0.77M (7.9K)\\
SITW      & 151 & 1006 & 0.50M (3K)\\ 
\quad low deg & 150 & 998 & 0.16M (735)\\
\quad high deg & 151 & 1003 & 0.20M (1.2K)\\ \hline

\end{tabular}}
\end{table}

\subsection{AMI}
The AMI Meeting corpus\footnote{\url{http://groups.inf.ed.ac.uk/ami/corpus/}} consists of over 100 hours of office meetings recorded in four different locations. The meetings are recorded using both close-talk and far-field  microphones, we use the former for diarization purpose. Since each speaker has their individual channels, we beamformed the audio into a single channel. We follow \cite{sun_2019,moni_clusterGAN} for splitting the sessions into the dev and eval partitions, ensuring that no speakers overlap between them. For our purposes, the AMI sessions represent audio collected in noise-free recording conditions.

\subsection{DIHARD}
The DIHARD speaker diarization challenges \cite{ryant2018first} were introduced in order to focus on hard diarization tasks, i.e., in-the-wild data collected with naturalistic background conditions. In this work, we use the development set from second DIHARD challenge. This corpus consists of data from multiple domains such as clinical interviews, audiobooks, broadcast news, etc. We make use of the 192 sessions in the single-channel task in this work. It is worth noting that a handful of sessions in this corpus contain only a single speaker.

\begin{table}[h]
\caption{Statistics of corpora used for speaker diarization}
\label{tab:corpora_diar}
\begin{tabular}{ccccc} \\ \hline
Corpus & \#Sessions & \#Spkrs/Session & \multicolumn{1}{c}{\begin{tabular}[c]{@{}c@{}}Session Duration\\ (min: ($\mu \pm \sigma$))\end{tabular}} \\ \hline
DIHARD & 192 & 3.48 & 7.44 $\pm$ 3.00\\
AMI (dev+eval) & 26 & 3.96 & 31.54 $\pm$ 9.06\\
ADOS-Mod3& 173 & 2 & 3.23 $\pm$ 1.50 \\ \hline
\end{tabular}
\end{table}

\subsection{ADOS-Mod3}
\label{subsec:ados}
One of the most challenging domains from the DIHARD evaluations included speech collected from children. Speaker diarization for these interactions involve additional complexities due to two reasons: (1) An intrinsic variability in child speech owing to developmental factors \cite{Lee1999Acousticsofchildrensspeech:,lee2014developmental}, and (2) Speech abnormalities due to underlying neuro-developmental disorder such as autism. To this end, we use 173 child-adult interactions consisting of excerpts from the administration of module 3 of the ADOS (Autism Diagnosis Observation Module) \cite{Lord2000}. These interactions involve children with sufficiently developed linguistic skills, i,e., ability to form complete sentences. All the children in this study had a diagnosis of autism spectrum disorder (ASD) or attention deficit hyperactivity disorder (ADHD). The sessions were collected from two different locations and manually annotated using the SALT transcription guidelines\footnote{\url{https://www.saltsoftware.com/media/wysiwyg/tranaids/TranConvSummary.pdf}}. Details of corpora used for speaker diarization is provided in Table \ref{tab:corpora_diar}.

%% file: sections/experiments.tex
\begin{figure}[h]
    \centering
    \includegraphics[width=\linewidth]{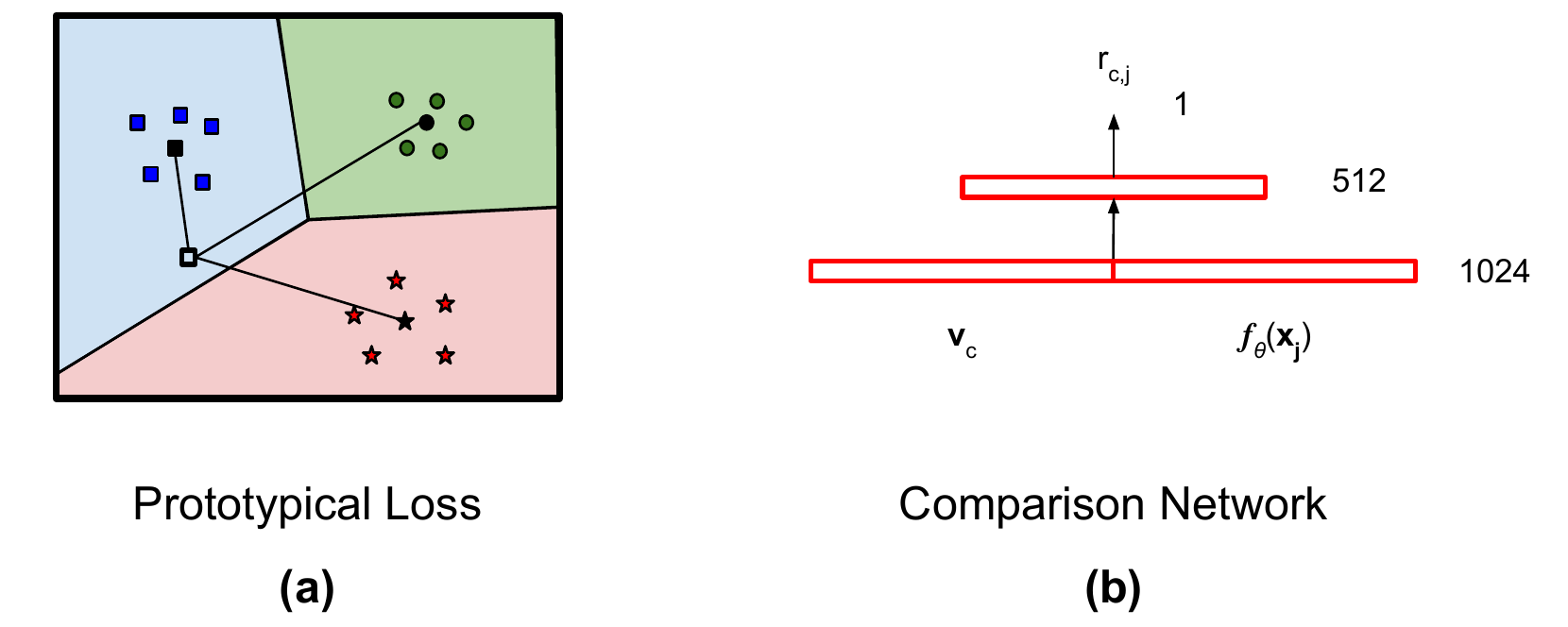}
    \caption{\textbf{(a)} Illustrating the training step in prototypical networks. Decision regions are indicated using background colors. For each class, prototypes are estimated as the centroid of supports (filled shapes). Given the query (unfilled shape), negative distances to each prototype are treated as logits. Adopted from \cite{koluguri2020}. (\textbf{b}) Comparison module in relation networks. The sum of support embeddings from class $c$ ($\mathbf{v}_c$) is concatenated with a query embedding ($f_\theta(\mathbf{x}_j)$) and input to the comparison network. $r_{c,j}$ is known as the relation score for query $\mathbf{x}_j$ with respect to class $c$ and treated as the logit.}
    \label{fig:metaLearningLoss}
\end{figure}

\section{Experiments and Results}
\label{sec:expts}

\subsection{Baseline Speaker Embeddings}
\label{subsec:exp_baseline}
In order to select a competitive and fair baseline to meta-learned embeddings, we first developed an implementation of x-vectors. 
Our model is similar to the Kaldi Voxceleb recipe\footnote{https://kaldi-asr.org/models/m7} with respect to training corpora and network architecture. We compare the reported performance of Kaldi embeddings with our implementation and select the best performing model as the baseline system.

As mentioned in Section \ref{subsec:data_vox}, we use the Vox2 and Vox1-dev corpora for embedding training. Similar to the Kaldi recipe, we extract 30-dimensional MFCC features using a frame width of 25ms and overlap of 15ms. We augment the training data with noise, music and babble speech using the MUSAN corpus\cite{snyder2015musan}, and reverberation using the RIR\_NOISES \footnote{\url{http://www.openslr.org/28}} corpus. The augmented data consist of 7323 speakers and 2.2M utterances. Following which, all utterances shorter than 4 seconds in duration and all speakers with fewer than 8 utterances each are removed to assist the training process. Cepstral mean normalization using a sliding window of 3 seconds was performed to remove any channel effects.

The model architecture consists of 5 time-delay layers which model temporal context information, followed by a statistical pooling layer to map into a utterance-level vector. This is followed by two feed-forward bottleneck layers with 512 units in each layer and the final layer which outputs speaker posterior probabilities. In contrast with the Kaldi implementation, we use Adam optimizer ($\beta_1$=0.9, $\beta_2$=0.99) to train the model, with an initial learning rate of 1e-3. The learning rate is increased to 2e-3 and progressively reduced to 1e-6. Dropout and batch normalization are used at all layers for regularization purpose. A minibatch of 32 samples is used at each iteration, while ensuring that utterances in each minibatch are of fixed duration to improve the training process. We accumulated gradients for every 4 minibatches before back propagation, which was observed to improve model convergence.

\begin{table}
\caption{Selecting a baseline system for speaker diarization. For each embedding and clustering method (AHC-f: AHC with fixed threshold, AHC-p: AHC with optimized threshold, bSC: binarized spectral clustering with normalized maximum eignegap), diarization error rate (DER \%) is provided for two settings: using oracle speaker count (Oracle) and estimated count (Est). }
\label{tab:spkrDiarBase}
\centering{
\begin{tabular}{ccp{6.1mm}p{6.1mm}p{6.1mm}p{6.1mm}p{6.1mm}p{6.1mm}} \hline
\multirow{2}{*}{Tool} & \multirow{2}{*}{Method} & \multicolumn{2}{c}{DIHARD} & \multicolumn{2}{c}{AMI} & \multicolumn{2}{c}{ADOSMod3} \\
 &  & Oracle & Est & Oracle & Est & Oracle & Est \\ \hline
 \rule{0pt}{2ex} 
\multirow{3}{*}{Kaldi} & AHC-f & \textbf{15.94} & 24.67 & 13.96 & 12.64 & 19.53 & 31.05 \\
 & AHC-o & - & 18.35 & - & 14.28 & - & 18.17 \\
 & bSC & 18.81 & 15.26 & 8.57 & 9.50 & 14.77 & 19.57 \\ \hdashline
 \rule{0pt}{2ex} 
\multirow{3}{*}{\begin{tabular}[c]{@{}c@{}}Ours\\ fc1\end{tabular}} & AHC-f & 17.09 & 24.47 & 15.40 & 14.49 & 18.82 & 33.14 \\
 & AHC-o & - & 18.74 & - & 14.55 & - & 20.18 \\
 & bSC & 18.81 & 14.62 & 7.95 & 14.51 & 15.85 & 21.37 \\ \hdashline
 \rule{0pt}{2ex} 
\multirow{3}{*}{\begin{tabular}[c]{@{}c@{}}Ours\\ fc2\end{tabular}} & AHC-f & 22.17 & 24.77 & 18.03 & 16.25 & 18.89 & 30.37 \\
 & AHC-o & - & 19.61 & - & 16.23 & - & 20.03 \\
 & bSC & 17.62 & \textbf{13.93} & \textbf{6.94} & \textbf{8.47} & \textbf{13.94} & \textbf{17.16} \\ \hline
\end{tabular}}
\end{table}

\subsection{Meta-learned embeddings}
\label{subsec:training_details}


We select DNN architectures for the meta-learning models similar to the baseline model in order to enable a fair comparison. 
We use the same network as x-vectors except for the final layer, i.e., we retain the time-delay layers, the stats pooling layer, and two fully connected layers with 512 units in each layer. 
The protonet model uses an additional two fully connected layers with 512 units in each layer. Embeddings extracted at the final layer are used for prototype computation and loss estimation. The relation network uses one additional fully connected layer (512 units) for the encoder network. The comparison network consists of three fully connected layers with 1024 units at the input, 512 units in the hidden layer and 1 unit at the output. 
For both networks, we use batch normalization which was observed to improve convergence. We do not use dropout in the meta-learned models following their respective original implementations \cite{snell2017prototypical,sung2018learning}. The number of trainable parameters for the baseline x-vector model, protonet and relation net (encoder + comparison) are 9.8M, 6.6M and 7.1M, respectively. 
We trained both protonets and relation nets using the Adam optimizer ($\beta_1$=0.9, $\beta_2$=0.99). The initial learning rate was set to 1e-4 and exponentially decreased ($\gamma$ = 0.9) every 10 episodes, where an episode corresponds to a single back-propagation step. The models were trained for 100K episodes with the stopping point determined based on convergence of smoothed loss function. The architecture and initialization strategies for all models are presented in Figure~\ref{fig:metaLearningArch}, while the meta-learning losses are illustrated in Figure~\ref{fig:metaLearningLoss}.

\textbf{Model Initialization:}
We use a part of the pre-trained x-vector model as an initialization for the meta-learning model. 
Specifically, we initialize the time-delay layers using the pre-trained weights from the corresponding layers from the x-vector model.
The fully connected layers are initialized uniformly at random between $[\frac{-1}{\sqrt{N}}, \frac{1}{\sqrt{N}}]$ where $N$ is the number of parameters in the layer. Empirically, we observed that the above initialization scheme provided a significant performance improvement in our experiments. 

\begin{figure*}[h]
\begin{tikzpicture}
  \node (img)  {\includegraphics[width=0.9\textwidth, height=7cm]{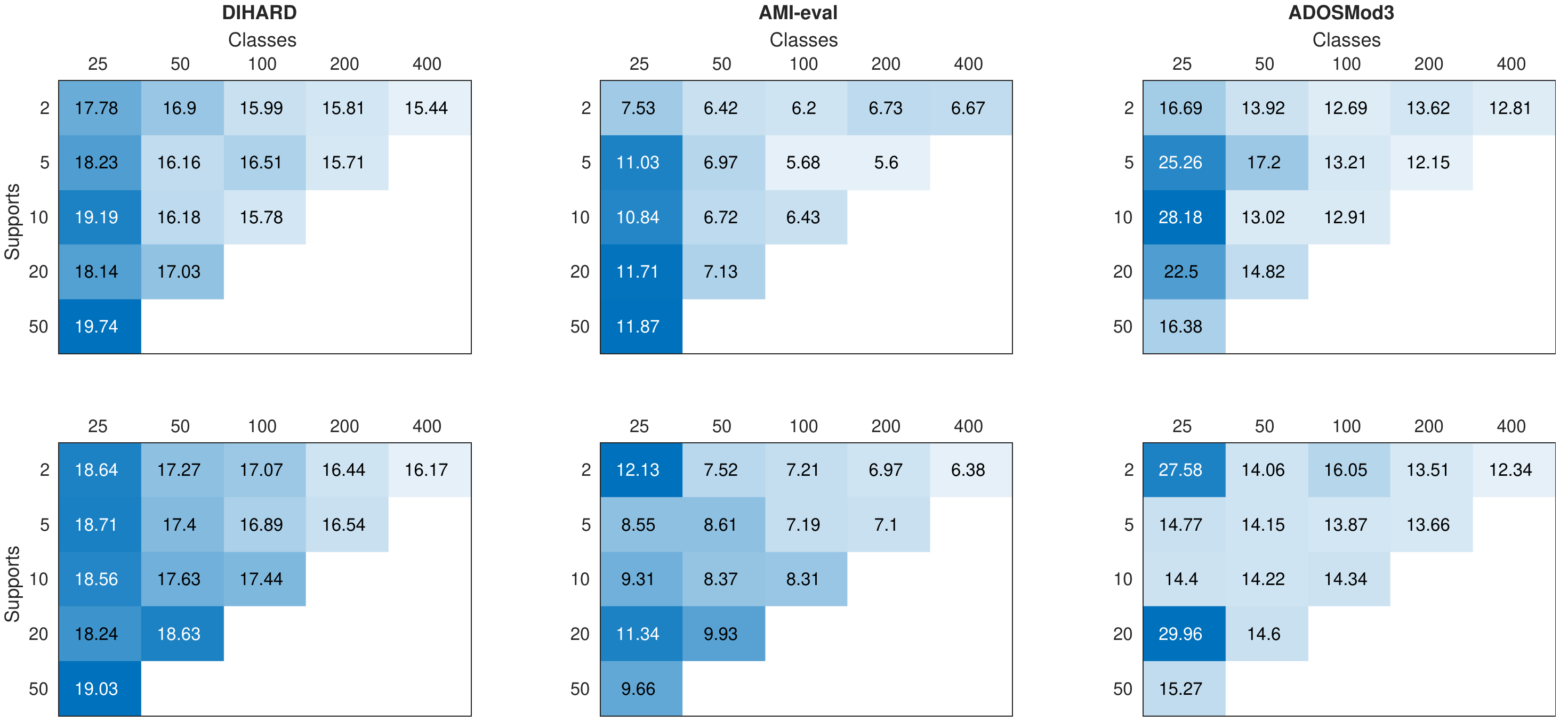}};
  \node[left=of img, node distance=0cm, rotate=90, anchor=center,yshift=-0.7cm] {\hspace{-1cm}  Relation Net \hskip 2.3cm Protonet};  
 \end{tikzpicture}
\caption{Speaker diarization performance (\% DER) across different corpora for different combinations of supports examples and training classes within an episode. Number of queries per class is always 1 in all experiments.}
\label{fig:wayShotExpt}
\end{figure*}

\begin{table}[h!]
\caption{Speaker diarization results comparing meta-learning models with x-vectors. x-vector+retrain represents mean DER computed with 3 trials}
\label{tab:spkrDiarCore}
\centering
\begin{tabular}{ccccccc} \\ \hline
\multirow{2}{*}{Method} & \multicolumn{2}{c}{DIHARD} & \multicolumn{2}{c}{AMI} & \multicolumn{2}{c}{ADOSMod3} \\
 & Oracle & Est & Oracle & Est & Oracle & Est \\ \hline
x-vectors & 17.62 & 13.93 & 6.94 & 8.47 & 13.94 & 17.16 \\
x-vector+retrain &  17.39 & 13.26 & 7.49 & 8.52 & 16.74 & 16.89 \\ \hdashline
\rule{0pt}{2ex}
Protonet & \textbf{15.44} & 12.96 & 6.67 & \textbf{7.31} & 12.81 & 17.22 \\ 
Relation Net & 16.17 & \textbf{12.65} & \textbf{6.38} & 8.94 & \textbf{12.34} & \textbf{16.19} \\ \hline
\end{tabular}
\end{table}

Since we borrow a part of the pre-trained x-vector model in our meta-learning models during initialization, we verify that any gains in performance obtained with meta-learning models do not arise from overtraining the x-vector model.
We conduct a sanity check experiment wherein we retrain the x-vector model similar to the meta-learning models. Specifically, we use the baseline model from Section \ref{subsec:exp_baseline} and retrain it using pre-trained weights for time-delay layers and random initialization for the fully-connected layers. The model was trained for 100K minibatches, which corresponds to the same number of episodes used for training meta-learning models.

\subsection{Speaker Diarization Results}
\label{subsec:spkrDiarResults}

We use the oracle speech activity detection for speaker diarization in order to study exclusively the speaker errors. We segment the session to be diarized into uniform segments 1.5 seconds long in duration and with an overlap of 0.75 seconds. Embedding clustering is performed using the NME-SC method as described in Section \ref{subsubsec:bSCNME}. During scoring, we do not use a collar similar to DIHARD evaluations. However, we discard speaker overlap regions since neither x-vectors nor meta-learned embeddings are trained to handle overlapping speech.

Table \ref{tab:spkrDiarBase} presents speaker diarization results for various baseline embeddings. We compare between pre-trained Kaldi embeddings, and both feed-forward bottleneck layers in our implementation. In addition to NME-SC for speaker clustering, we use AHC on PLDA scores using two methods for estimating number of speakers: (1) A fixed threshold parameter of 0, (2) Tuned threshold parameter using a development set. We tuned the parameter using two-fold cross validation for DIHARD and ADOS-Mod3, and the AMI-dev set for the AMI corpus.

\begin{figure*}[h]
    \centering
    \includegraphics[width=\textwidth]{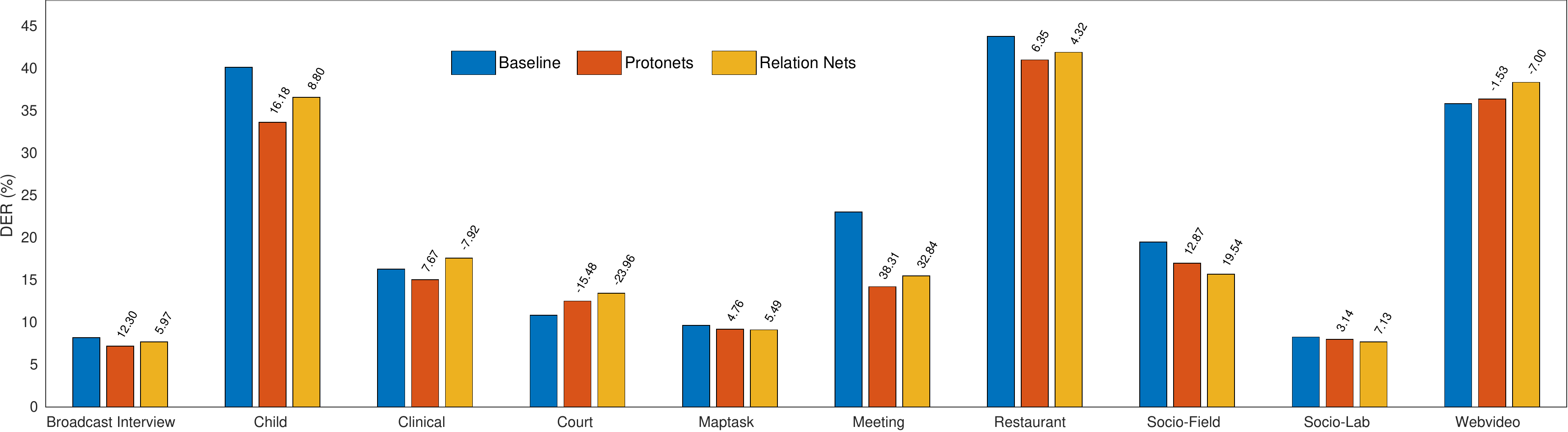}
    \caption{Diarization performance across domains in DIHARD. For each domain, the mean DER across sessions is provided for baseline (x-vectors), protonets and relation nets. The relative change in DER (\%) with respect to the baseline is given next to the bar (postive: DER reduction)}
    \label{tab:domain_results}
\end{figure*}

First, we notice that AHC is quite sensitive to the threshold parameter when estimating the number of speakers across all corpora and clustering methods. DER reduction using a fine-tuned threshold is particularly significant for the ADOS-Mod3 corpus with nearly 13\% absolute improvement for fc1, and 10\% for fc2 embeddings extracted using our network. 
In some cases on the DIHARD and AMI corpora the DER obtained by fine-tuning the threshold is lower than when oracle number of speakers is used, similar to observations in \cite{pal_clusterGAN2020}.
Next, fc1 embeddings outperform fc2 embeddings when clustering using AHC and PLDA scores, consistent with findings from \cite{snyder_xvec2017}. However, when cosine affinities are used with NME-SC we notice that the layer closer to the cross-entropy objective (fc2) results in a lower DER. This is the case both when oracle number of speakers are used as well as when they are estimated using the maximum eigengap value. The combination of fc2 embeddings with NME-SC method returns the lowest DERs for most conditions. Further, NME-SC removes the need for a separate development set for estimating the threshold parameter. Hence, we adopt this as the diarization baseline method in all our experiments.

In Table \ref{tab:spkrDiarCore}, we compare the baseline with the meta-learning models. \textit{x-vector+retrain} represents mean results from 3 trials of the sanity check experiment described in the Section \ref{subsec:training_details}. Both meta-learning models were trained for 100K episodes. Within each episode, 400 classes were randomly chosen without replacement from the training corpus. Following which, 3 samples were chosen without replacement from each class. Two samples were treated as supports, while the third sample was treated as query. 
From the results, we note that retraining the x-vector model provides minor DER improvement on the DIHARD corpus while performance worsens on the AMI corpus. The meta-learning models outperform the baselines in most cases, although improvements depend on the corpus and setting. On the DIHARD corpus consisting of challenging domains, protonets result in 12.37\% relative improvement given oracle number of speakers and 6.96 \% improvement when the number of speakers are estimated. Relation networks show a slight degradation when compared to protonets. This difference is more on a relatively clean corpus such as AMI while estimating number of speakers. 
In the following experiments, we analyze which setups contribute to improvements in performance over x-vectors.

\subsubsection{Effect of classes within a task}

While training meta-learning models, previous works \cite{snell2017prototypical, sung2018learning, vinyals2016matching} often carefully control the number of classes (\textit{way}) within an episode and the number of supports per class (\textit{shot}) so as to match the evaluation scenario. 
Drawing analogies with speaker diarization, a typical session consists of $\mathcal{O}(1)$ speakers (\textit{way}), with $\mathcal{O}(10)$ utterances per speaker (\textit{shot}). In this experiment we vary hyper-parameters for both protonets and relation nets, and study the effect on DER. We vary the \textit{way} and \textit{shot} between 25 to 400, and 2 to 50, respectively, and train a new meta-learning model for each configuration. Results are presented in Fig~\ref{fig:wayShotExpt}. 

A common effect across different corpora and models is that the number of speakers (classes) is an important parameter for diarization performance. Increasing the number of speakers in an episode favours DER. This is similar to previous findings in few-shot image recognition \cite{snell2017prototypical}, where during training, a higher \textit{way} than expected during testing was found to provide the best results. However, the effect of supports per class on DER is not straightforward. When a large number of classes is used, increasing supports provides little to no improvements in both protonets and relation nets. Upon reducing the number of classes, the performance degrades with more supports across most models. This suggests a possibility of over-fitting due to large number of supports even though the configuration closely resembles a test session. It is more beneficial to increase the number of classes within an episode during training.



\subsubsection{Performance across different domains in DIHARD}

It is often useful to understand the effect of conversation type, including speaker count, spontaneous speech and recording setups on the diarization performance. We study this using the domain labels \cite{ryant2019second} available for the DIHARD corpus. For each domain, we compute the mean DER across sessions using the baseline model as well as the meta-learning models. Oracle speaker count is used during clustering in order to exclusively study the effect of domain factors. We do not include the Audiobooks domain in this experiment since all the models return the same performance on account of sessions consisting of only one speaker.
We present the results in \mbox{Table \ref{tab:domain_results}}.

We note that there exists considerable variation between domains in terms of the DER improvement between x-vectors and meta-learning models. Broadcast news, child, maptask, meeting and socio-field domains show significant gains due to meta-learning models. Specifically, meeting and child domains benefit upto 38.31 \% and 16.18 \% relative DER improvement from protonets. Diarization in the court domain degrades in performance consistently between protonets and relation nets, with up to 20.05 \% relative degradation for relation networks. Upon a closer look at the court and meeting domains to understand this difference, we note that both domains contain similar number of speakers per session (Court: 7, Meeting: 5.3). However, the domains differ in the data collection setup: court sessions are collected by averaging audio streams from individual table-mounted microphones, while meeting sessions are collected using a single table microphone distant from all the participants \cite{ryant2019second}. Among the socio-linguistic interview domains, interviews recorded in the field under diverse locations and subject age groups (socio-field) result in a larger DER improvement over those collected under quiet conditions (socio-lab). Socio-lab contains recording from both close-talking and distant microphones, hence it is not immediately clear whether microphone placement alone is a factor in DER improvement. 
Child and restaurant domains show variation in DER reduction although they perform similar with the baseline models, suggesting that background noise types affect benefits from meta-learning.
Overall, most domains that include in-the-wild data collection show improvements with meta-learning.

\subsubsection{Performance across different child age groups}

As mentioned in Section \ref{subsec:ados}, 
automatic child speech processing has been considered a hard problem when compared to processing adult speech.
More recently, the child domain returned one of the highest DERs during the DIHARD evaluations \cite{Xie2019}, illustrating the challenges of working with child speech for diarization.
Considering meta-learning models return significant improvement over x-vectors for child domain, we attempt to understand gains in DER by controlling for the age of the child. Children develop linguistic skills as they grow up, hence child age is a reasonable proxy for their linguistic development. 
We select sessions from the ADOS-Mod3 corpus where we have access to the child age metadata.
We compute the DER for each child using the respective baseline and meta-learned models described in Section~\ref{subsec:training_details}. For children where two sessions are available, we compute the mean DER per child. We study the effect of child age on DER by grouping child age into 3 groups with approximately equal number of children in each set. Children below 7.5 years of age are collected in the Low age group, children between 7.5 years and 9.5 years of age are collected in the Mid age group, and children above 9.5 years of age are collected in the High age group.

\begin{table}[h]
\caption{Analysis of child-adult diarization performance on the ADOS-Mod3 corpus. For each age group, mean DER (\%) of sessions in each group are presented along with relative improvement in parenthesis.}
\label{tab:age_expt}
\centering
\begin{tabular}{cccc} \\ \hline
Model & Low & Mid & High \\ \hline
\rule{0pt}{3ex} Baseline & 17.36 & 13.42 & 13.77 \\
\rule{0pt}{3ex} Protonet & 15.77 (9.16) & \textbf{12.39 (7.68)} & 12.33 (10.46)\\
\rule{0pt}{3ex} Relation Net & \textbf{15.69 (9.62)} & 12.82 (4.47) & \textbf{11.37 (17.43)} \\ \hline
\end{tabular}
\end{table}

From the results in Table~\ref{tab:age_expt}, we notice that the Low age group returns the highest DER, while Mid and High age groups return similar performance across models. Given that children in the Low age group are more likely to exhibit speech abnormalities, this result illustrates the relative difficulty in automatic speech processing under such conditions. Improvements in DER from meta-learning models are distributed across all age groups.
A consistent improvement of 10\% relative DER among the Low age group is particularly encouraging given the challenging nature of such sessions. The high age group exhibits similar improvements in DER, with the relation networks providing upto 17.43 \% relative gains. 

\subsection{Speaker Verification Results}

We use speaker verification as another application task to illustrate the generalized speaker information captured by meta-learned embeddings. Similar to speaker diarization, we first evaluate our implementation of the baseline with the pre-trained Kaldi embeddings. We use the test partition of Voxceleb corpus, the eval set in VOiCES corpus and the eval set in SITW corpus in our experiments.
We use the core-core condition in the SITW corpus where a single speaker is present in both utterances during a trial. For all models, we score trials using PLDA after performing dimension reduction to 200 using LDA and length-normalization. The PLDA model is trained using the same data for embedding training, i.e., Vox2 corpus and the dev set of Vox1 corpus. Speakers in the SITW corpus which overlap with the Voxceleb corpus were removed from the trials before evaluation.
We use equal error rate (EER) as the metric to select the best performing baseline system. 
Since cosine scoring returned significantly high EERs relative to PLDA, we did not investigate it further.
Results are provided in Table~\ref{tab:base_sv}.

\begin{table}[h]
\caption{Selecting a baseline system for speaker verification. Results are presented as equal error rate (EER \%)}
\label{tab:base_sv}
\centering{
\begin{tabular}{cccc} \hline
Embedding & Vox1-test & VOiCES & SITW \\ \hline
Kaldi & 3.128 & 10.300 & 4.054 \\
Ours:fc1 & \textbf{2.815} & \textbf{8.591} & \textbf{3.856} \\
Ours:fc2 & 3.006 & 9.854 & 4.087 \\ \hline
\end{tabular}}
\end{table}

We notice that embeddings from both layers in our implementation outperform or closely match the Kaldi implementation. Similar to observations from Section \ref{subsec:exp_baseline} and \cite{snyder_xvec2017} fc1 embeddings fare better than fc2 embeddings when scored with PLDA. We select fc1 embeddings as the baseline speaker verification method.


\begin{table}[h]
\caption{Speaker verification results comparing meta-learning models with x-vectors. Results presented using EER and minDCF computed at $P_{target} = 0.01$ }
\label{tab:spkrVerCore}
\begin{tabular}{ccccccc} \\ \hline
\multirow{2}{*}{Model} & \multicolumn{2}{c}{Vox1-test} & \multicolumn{2}{c}{VOiCES} & \multicolumn{2}{c}{SITW} \\
 & EER & DCF & EER & DCF & EER & DCF  \\ \hline
\rule{0pt}{3ex} Baseline & \textbf{2.815} & 0.311 & 8.591 & 0.696 & 3.856 & 0.359 \\
\rule{0pt}{3ex} Protonets & 2.831 & \textbf{0.299} & \textbf{7.837} & \textbf{0.646} & \textbf{3.560} & \textbf{0.347} \\ 
\rule{0pt}{3ex} Relation Net & 2.884 & 0.313 & 8.238 & 0.690 & 3.725 & 0.370 \\ \hline
\end{tabular}
\end{table}

When comparing meta-learning models, we use the same models developed in Section~\ref{subsec:spkrDiarResults}. In addition to EER, we present results using the minimum detection cost function (minDCF) computed at $P_{target} = 0.01$. From Table~\ref{tab:spkrDiarCore}, we note that meta-learning models outperform x-vectors in most settings except in the case of Voxceleb corpus when EER is used. Both protonets and relation nets return similar EER and minDCF for the Voxceleb corpus. Interestingly, we achieve notable improvements on the relatively more challenging corpora. Protonets provide up to 8.78\% and 7.68\% EER improvements in the VOiCES and SITW corpora, respectively, with similar improvements in minDCF. While relation nets provide better performance than x-vectors in the above corpora, they do not outperform protonets in any setting. This suggests that using a predefined distance function (namely squared Euclidean in protonets) might be beneficial overall when compared to learning a distance metric using relation networks for speaker verification application.

\begin{table*}[h]
\caption{Analysis of speaker verification based on microphone location (Near: Near-field, Far: Far-field, Obs: Fully obscured) in VOiCES corpus and level of degradation artefacts in SITW corpus}
\label{tab:spkrVerRobust}
\centering
\begin{tabular}{cccccccccccc} \\ \hline
 & \multicolumn{6}{c}{VOiCES (mic location)} &  & \multicolumn{4}{c}{SITW (degradation level)} \\ \cline{2-7} \cline{9-12}
\rule{0pt}{2ex} 
Model & \multicolumn{2}{c}{Near} & \multicolumn{2}{c}{Far} & \multicolumn{2}{c}{Obs} &  & \multicolumn{2}{c}{Low} & \multicolumn{2}{c}{High} \\
 & EER & DCF & EER & DCF & EER & DCF &  & EER & DCF & EER & DCF \\ \hline
\rule{0pt}{3ex} Baseline & 3.907 & \textbf{0.3407} & 7.311 & \textbf{0.5797} & 22.65 & 0.9375 &  & \textbf{3.401} & 0.3463 & 4.815 & 0.445 \\
\rule{0pt}{3ex} Protonets & \textbf{3.801} & 0.376 & \textbf{7.132} & 0.6337 & \textbf{20.58} & \textbf{0.9366} & & 3.537 & \textbf{0.3281} & 4.414 & \textbf{0.4268} \\
\rule{0pt}{3ex} Relation Net & 3.872 & 0.3521 & 7.618 & 0.6282 & 21.24 & 0.9527 &  & 3.81 & 0.3467 & \textbf{4.414} & 0.4525 \\ \hline
\end{tabular}
\end{table*}

\subsubsection{Robust Speaker Verification}
Since VOiCES and SITW corpora return the most improvement for speaker verification, we take a closer look at which factors benefit meta-learning. For each corpus, we make use of annotations for the microphone location and channel degradation to create new trials for speaker verification. 

\begin{figure}[h]
    \centering
    \includegraphics[scale=0.27]{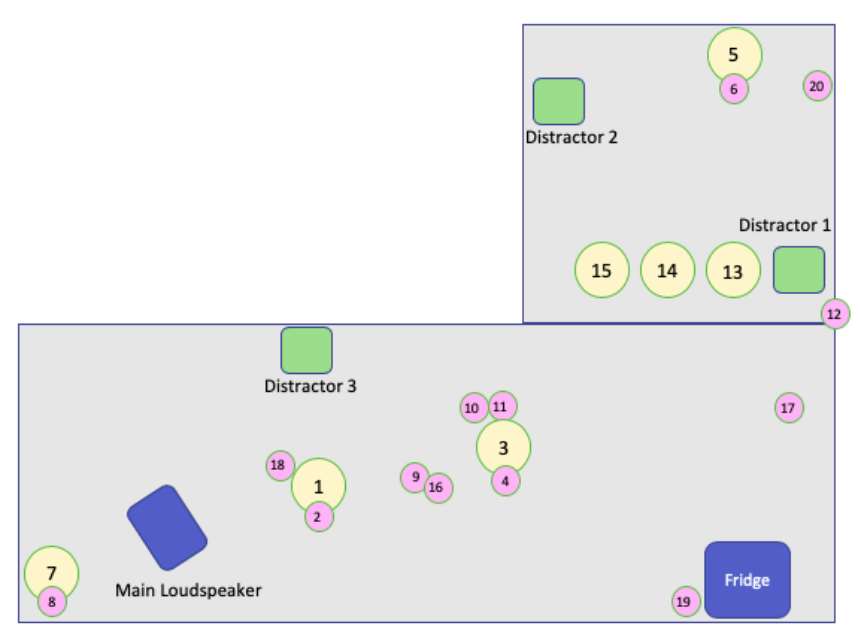}
    \caption[An example room configuration from the VOiCES corpus]{An example room configuration from the VOiCES corpus\protect\footnotemark. Microphones are represented using circles. }
    \label{fig:voicesRoom}
\end{figure}

In the VOiCES corpus, we collect playback recordings from rooms 3 and 4 present in the eval subset. Within these recordings, we distinguish between the utterances based on the microphone placement with respect to the loudspeaker (audio source). Specifically, we create three categories: (1) utterances collected using mic1 and mic18 are treated as near-field, being closest to the source, (2) utterances collected from mic19 are treated as far-field, and (3) utterances collected from mic12 are treated as obscured, since they are fully obscured by the wall. 
While creating the trials for each category, we ensure that the ratio of target to nontarget pairs remain approximately equal to the overall eval set trial. An example room configuration is presented in Figure~\ref{fig:voicesRoom}.

\footnotetext{Figure adapted from https://voices18.github.io/rooms/}

From the SITW corpus, we use the metadata annotations for level of degradation. The corpus includes multiple degradation artifacts: reverberation, noise, compression, etc, among others. The level of degradation for the most prominent artefact was annotated manually on a scale of 0 (least) to 4 (maximum). We use the trials available as part of the eval set which are annotated with the degradation level. We group the trials into two levels: low (deg0 and deg1) and high (deg3 and deg4). Note that the utterances contain multiple types of degradation in each level. 
Details of target and imposter pairs for SITW corpus (degradation level) and VOiCES corpus (microphone placement) are present in Table~\ref{tab:spkrVerDataStats}. Speaker verification results using EER and minDCF are presented in Table~\ref{tab:spkrVerRobust}.

We notice that no single model performs the best across multiple conditions. When controlled for microphone placement in VOiCES, protonets return the best EER at all locations. The margin of improvement remains approximately the same when only the distance from source is considered: 2.71\% for near-field and 2.45\% for far-field. The margin improves to 9.14\% when the microphone is fully obscured by a wall and placed close to distractor noises. Interestingly, these improvements are not reflected in the minDCF scores in the absence of noise, where x-vectors outperform both meta-learning models. 
We believe that improvements in EER and minDCF in VOiCES corpus primarily arise from utterances collected in obstructed locations and in close vicinity of distractor noises. The experiments in SITW corpus focus on the strength of such noise conditions. Under low degradation levels, we see that x-vectors return the least EER, although their performance is not consistent with minDCF. Meta-learning models continue to work better in higher degradation levels, providing 8.3\% reduction in 4.1\% reduction in EER and minDC, respectively. 



%% file: sections/conclusion.tex
\section{Conclusions}

We proposed neural speaker embeddings trained with the meta-learning paradigm, and evaluated on corpora representing different tasks and settings. 
In contrast to conventional speaker embedding training which optimizes on a single classification task, we simulate multiple tasks by sampling speakers during the training process. 
Meta-learning optimizes on a new task at every training iteration, thus improving generalizability to an unseen task.
We evaluate two variants of meta-learning, namely prototypical networks and relation networks on speaker diarization and speaker verification.
We analyze the performance of meta-learned speaker embeddings in challenging settings such as far-field recordings, child speech, fully obstructed microphone collection and in the presence of high noise degradation levels.
The results indicate the potential of meta-learning as a framework for training multi-purpose speaker embeddings.


In the future, we plan to investigate combining clustering objectives such as deep clustering \cite{hershey_dpcl2016, pmlr-v70-law17a} with meta-learning.
A combination of protonets and relation networks with similar metric learning approaches such as matching networks and induction networks will also be explored to study complementary information between them. Further generalization to unseen classes can be obtained by incorporating domain adversarial learning techniques with the meta-learning paradigm.